\documentclass{aa}
\usepackage{graphicx}
\usepackage{txfonts}
\begin{document}
\title{
Comparison of dynamical model atmospheres of Mira variables with 
mid-infrared interferometric and spectroscopic observations
}


\author{K.~Ohnaka\inst{1}, 
M.~Scholz\inst{2,3}, 
P.~R.~Wood\inst{4}
}

\offprints{K.~Ohnaka}

\institute{
Max-Planck-Institut f\"{u}r Radioastronomie, 
Auf dem H\"{u}gel 69, 53121 Bonn, Germany\\
\email{kohnaka@mpifr-bonn.mpg.de}
\and
Institut f\"ur Theoretische Astrophysik der Universit\"at Heidelberg, 
Albert-Ueberle-Str. 2, 69120 Heidelberg, Germany
\and
School of Physics, University of Sydney, Sydney, NSW 2006, 
Australia
\and
Research School for Astronomy and Astrophysics, Australian National 
University, Canberra, ACT 2600, Australia
}

\date{Received / Accepted }

\abstract{
We present a comparison of dynamical model atmospheres 
with mid-infrared ($\sim$11~\mbox{$\mu$m}) interferometric and 
spectroscopic observations of the Mira variable \mbox{$o$~Cet}.  
The dynamical model atmospheres of Mira variables pulsating 
in the fundamental mode can fairly 
explain, without assuming ad-hoc components, 
the seemingly contradictory mid-infrared spectroscopic 
and interferometric observations of \mbox{$o$~Cet}: the 11~\mbox{$\mu$m}\ 
sizes measured in the bandpass without any salient spectral 
features are about twice as large as those measured in the 
near-infrared.  Our calculations of synthetic spectra 
show that the strong absorption due to 
a number of optically thick \mbox{H$_2$O}\ lines is filled in 
by the emission of these \mbox{H$_2$O}\ lines originating in the 
geometrically extended layers, providing a possible physical 
explanation for the picture proposed by Ohnaka (\cite{ohnaka04a}) 
based on a semi-empirical modeling.  
This filling-in effect results in rather featureless, 
continuum-like spectra in rough agreement with the observed 
high-resolution 11~\mbox{$\mu$m}\ spectra, although 
the models still predict the \mbox{H$_2$O}\ lines to be more pronounced 
than the observations.  
The inverse P-Cyg profiles of some strong \mbox{H$_2$O}\ lines 
observed in the 11~\mbox{$\mu$m}\ spectra can also be reasonably 
reproduced by our dynamical model atmospheres.  
The presence of the extended \mbox{H$_2$O}\ layers manifests itself as 
mid-infrared angular diameters much larger than the continuum 
diameter.   The 11~\mbox{$\mu$m}\ uniform-disk diameters 
predicted by our dynamical model atmospheres are in fair agreement 
with those observed with the Infrared Spatial Interferometer (ISI), 
but still somewhat smaller than the observed diameters.  
We discuss possible reasons for this discrepancy and 
problems with the current dynamical model atmospheres of Mira 
variables.  
\keywords{infrared: stars -- techniques: interferometric -- 
stars: atmospheres -- stars: circumstellar matter -- 
stars: AGB and post-AGB -- stars: individual: $o$~Cet}
}   

\titlerunning{Dynamical model atmospheres of Miras}
\maketitle

\section{Introduction}
\label{sect_intro}
Mira variables exhibit large-amplitude pulsation, which leads 
to remarkable temporal variations of brightness, effective temperature, 
and geometrical extension of the atmosphere.  A better 
understanding of the structure of such dynamical atmospheres 
is crucial for unraveling the mass loss mechanism in these 
stars.  
Recent interferometric observations in the near-infrared as well 
as in the mid-infrared have revealed that the angular diameters of
Mira variables increase toward longer wavelengths.  Mennesson et
al. (\cite{mennesson02}) show that the \mbox{$L^{\prime}$}-band diameters 
of Mira variables are 20--100\% larger than those measured in the 
\mbox{$K^{\prime}$}\ band.  Millan-Gabet et al. (\cite{millan-gabet05}) 
carried out angular diameter measurements in the $J$, $H$, and 
$K$ bands for a number of Mira variables and show the general 
trend: $J$-band diameter $<$ $H$-band diameter $<$ $K$-band diameter. 
This trend extends into the mid-infrared, as demonstrated by the 
11~\mbox{$\mu$m}\ observations of Weiner et al. (\cite{weiner00}, 
\cite{weiner03a}, and \cite{weiner03b}, hereafter W00, WHT03a, 
WHT03b, respectively) using ISI.  They found that the 11~\mbox{$\mu$m}\ diameters 
of the Mira variables \mbox{$o$~Cet}, \mbox{R~Leo}, and \mbox{$\chi$~Cyg}\ are approximately 
twice as large as those observed in the $K$ band.  
In the optical, on the other hand, the angular sizes of Mira 
variables increase systematically toward shorter wavelengths, 
from 1~\mbox{$\mu$m}\ down to 0.65~\mbox{$\mu$m}, 
as Ireland et al. (\cite{ireland04a}) have recently shown.   

Such wavelength dependence of the angular sizes of 
Mira variables has often been attributed to dust shells, 
whose flux contribution due to dust thermal emission increases from 
the near-infrared to the mid-infrared.  In fact, the inner region of 
dust shells around Mira variables has been studied using 
mid-infrared interferometric data 
(e.g., Danchi et al. \cite{danchi94}; 
Tevousjan et al. \cite{tevousjan04}; Ohnaka et al. \cite{ohnaka05}).  
The increase of the angular sizes toward shorter wavelengths 
in the optical can also be attributed to the increase of 
scattered light due to dust grains (e.g., Ireland et al. 
\cite{ireland04a}).  Optical interferometric polarimetry 
of Mira variables carried out by Ireland et al. (\cite{ireland05}) 
also suggests that the scattering due to dust grains is 
responsible for the systematic increase of the angular size 
shortward of $\sim$1~\mbox{$\mu$m}.  

Nevertheless, the increase of the angular sizes of Mira variables 
from the near-infrared to the mid-infrared revealed by the 11~\mbox{$\mu$m}\ 
ISI observations of W00, WHT03a, and WHT03b cannot be solely attributed 
to the presence of extended dust shells for the following reason.  
The presence of an extended dust shell well separated from the 
underlying atmosphere 
leads to a steep drop of visibilities that appears 
only at low spatial frequencies, 
and the effect of such an extended dust shell is to 
lower the total visibility by an amount equal to the fractional 
flux contribution of the dust shell at the wavelength at issue.  
This effect is already taken into account in the derivation of 
the 11~\mbox{$\mu$m}\ uniform-disk diameters by W00, WHT03a, and WHT03b.  
Still, the angular sizes obtained at 11~\mbox{$\mu$m}\ are by a factor 
of $\sim$2 larger than those measured in the near-infrared.  

On the other hand, there is mounting evidence suggesting that gaseous 
layers extending to a few stellar radii are responsible for the
increase of the angular size from the near-infrared to the mid-infrared.  
Mennesson et al. (\cite{mennesson02}) as well as Schuller et al. 
(\cite{schuller04}) suggested that the 
increase of the angular sizes of Mira variables from the $K$ band 
to the \mbox{$L^{\prime}$}\ band may be attributed to extended gaseous layers.  
Perrin et al. (\cite{perrin04}) attempted to interpret the 
visibilities of Mira variables observed in the near-infrared as well 
as in the 11~\mbox{$\mu$m}\ region, using a model of a molecular shell 
where the molecular opacity and its wavelength dependence 
were treated as free 
parameters rather than calculated from detailed line lists.  
They conclude 
that the observed near-infrared and 11~\mbox{$\mu$m}\ visibilities 
can be reproduced by shells extending to $\sim$2.2~\mbox{$R_{\star}$}\ with 
temperatures of 1500--2000~K and optical depths of 0.2--1.2 at 
2.03~\mbox{$\mu$m}.   Weiner (\cite{weiner04}) used a similar model 
but with molecular opacities calculated from the detailed 
\mbox{H$_2$O}\ line list of Partridge \& Schwenke (\cite{partridge97}, 
hereafter PS97) and reached a similar conclusion 
based on the model fitting to the near-infrared and 11~\mbox{$\mu$m}\ 
visibilities of \mbox{$o$~Cet}.

Then, the high-resolution 11~\mbox{$\mu$m}\ spectra of Mira variables 
presented by WHT03a and WHT03b have appeared as a puzzling 
piece to this picture.  
These spectra obtained using the TEXES instrument with 
a spectral resolution of $\sim \! \! \! 10^{5}$ (Lacy et
al. \cite{lacy02}) 
show that no salient spectral feature is present within the 
narrow bandpasses used in the ISI observations, and 
this absence of spectral features  
appears to contradict the interpretation that the molecular 
layers close to the star are responsible for the increase of 
the angular size from the near-infrared to the mid-infrared.  
While the 11~\mbox{$\mu$m}\ spectra were not calculated 
or not well reproduced 
by Perrin et al. (\cite{perrin04}) or Weiner (\cite{weiner04}), 
Ohnaka (\cite{ohnaka04a}) demonstrates that the angular diameters and 
high-resolution spectra of Mira variables 
obtained in the 11~\mbox{$\mu$m}\ region can be simultaneously explained 
by an optically thick, dense warm water vapor envelope.  
Using a two-layer model for this warm water vapor envelope, 
Ohnaka (\cite{ohnaka04a}) found that the absorption due to the dense 
water vapor is 
filled in by emission from the extended part of the water 
vapor envelope, making the spectra almost featureless.  
On the other hand, the presence 
of the water vapor envelope manifests itself as an increase of the 
angular diameter from the near-infrared to the mid-infrared.  

While such ad-hoc models do not address the question about the 
physical processes responsible for the formation of the warm 
water vapor envelope, they are useful for illustrating 
the basic picture of the outer atmosphere and deriving its 
approximate physical properties, providing us with 
insights on possible next steps.  
In fact, Ohnaka (\cite{ohnaka04a}) shows that 
the intensity profiles predicted by the two-layer 
model for the warm water vapor envelope resemble those 
predicted by the dynamical models studied by Tej et al. (\cite{tej03a}), 
suggesting the possibility that the large-amplitude pulsation 
in Mira variables may be responsible for the existence of the warm 
water vapor envelope.  

In the present paper, we examine this possibility by comparing 
the observed 11~\mbox{$\mu$m}\ diameter and the high-resolution 
spectra of \mbox{$o$~Cet}\ with those predicted by dynamical  
model atmospheres.  

\section{Dynamical model atmospheres}

\begin{table}
\caption {Summary of dynamical models used in the present work. 
\mbox{$R_{1.04}$}\ is the radius defined by the position of the layer where the
optical depth in the 1.04~\mbox{$\mu$m}\ continuum is equal to unity.  
\mbox{$R_{\rm p}$}\ is the Rosseland radius of the parent star, which is an 
initial, static model.  The P models have \mbox{$R_{\rm p}$}\ = 241~\mbox{$R_{\sun}$}, 
while the M models have \mbox{$R_{\rm p}$}\ = 260~\mbox{$R_{\sun}$}.  
\mbox{$T_{\rm eff}$}\ is defined as 
$(L/4 \pi \sigma R_{\rm Ross}^2)^{1/4}$, where $R_{\rm Ross}$ is 
the Rosseland radius of a dynamical model, 
and $\sigma$ is the Stefan-Boltzmann constant. 
$T_{\rm 1.04} = (L/4 \pi \sigma R_{1.04}^2)^{1/4}$ gives a 
representative temperature of the continuum-forming layers.  
}
\begin{tabular}{l l l l l l}\hline
Model & Cycle+Phase & \mbox{$R_{1.04}$}/\mbox{$R_{\rm p}$} & \mbox{$T_{\rm eff}$}\ (K) & 
$T_{\rm 1.04}$ (K) & $L$ (\mbox{$L_{\sun}$}) \\ 
\hline
P05   & 0+0.5  & 0.90 & 2160 & 2500 & 1650 \\
P12   & 1+0.23 & 1.30 & 2610 & 2680 & 4540 \\
P13n  & 1+0.30 & 1.26 & 2310 & 2530 & 3450 \\
P14n  & 1+0.40 & 1.19 & 2080 & 2510 & 2920 \\
P15n  & 1+0.50 & 0.84 & 1800 & 2690 & 1910 \\
P22   & 2+0.18 & 1.26 & 2640 & 2700 & 4400 \\
P23n  & 2+0.30 & 1.24 & 2470 & 2590 & 3570 \\
P24n  & 2+0.40 & 1.16 & 2210 & 2420 & 2380 \\
P25   & 2+0.53 & 0.91 & 2200 & 2500 & 1680 \\
P35   & 3+0.5  & 0.81 & 2270 & 2680 & 1760 \\
\hline
M05   & 0+0.49 & 0.84 & 2310 & 2420 & 1470 \\
M12n  & 1+0.21 & 1.18 & 2410 & 2540 & 3470 \\
M14n  & 1+0.40 & 0.91 & 2110 & 2400 & 1670 \\
M15   & 1+0.48 & 0.83 & 2460 & 2530 & 1720 \\
M22   & 2+0.25 & 1.10 & 2330 & 2490 & 2850 \\
M23n  & 2+0.30 & 1.03 & 2230 & 2460 & 2350 \\
M24n  & 2+0.40 & 0.87 & 2160 & 2410 & 1540 \\
M25n  & 2+0.50 & 0.79 & 2770 & 2780 & 2250 \\
\hline

\label{table_model}
\end{tabular}
\end{table}

Since the details of the dynamical model atmospheres used in the
present work are described in Hofmann et al. (\cite{hofmann98}), 
Tej et al. (\cite{tej03b}), Ireland et al. (\cite{ireland04b}), 
and Ireland et al. (\cite{ireland04c}) (hereafter, HSW98, 
TLSW03, ISW04, and ISTW04, respectively), 
we only summarize the properties of the models, in particular those 
relevant to our calculations.  
These dynamical model atmospheres are specified by 
the period, stellar mass, luminosity, and Rosseland radius of the 
static parent star, which is an initial model before 
the star is brought into self-excited pulsation. 
In the present work, we use the P and M series, 
as given in Table~\ref{table_model}.  
Both P and M models are fundamental-mode pulsators with a period of
332~days, a parent-star luminosity of 
3470~\mbox{$L_{\sun}$}, and masses of 1.0~\mbox{$M_{\sun}$}\ (P series) and 1.2~\mbox{$M_{\sun}$}\ 
(M series).  
Here, 332~days is the formal period of the parent star, whereas the
actual periods of the dynamical models are slightly different 
(P models: 341~days; M models: 317~days; \mbox{$o$~Cet}: 334~days; 
see ISTW04). 
Throughout the present work, we define the continuum radius by 
the position of the layer where the optical depth measured in the
continuum at 1.04~\mbox{$\mu$m}\ is equal to unity.  
The Rosseland radius, which is also used as a ``stellar radius'' in 
some works, is often close to the 1.04~\mbox{$\mu$m}\ continuum radius.  
It should be mentioned, however, that it is not always the case, 
as discussed in ISW04 and ISTW04. 

Comparison of dynamical model atmospheres with various 
observations implies that Mira variables are fundamental-mode 
pulsators.  
Woodruff et al. (\cite{woodruff04}) have recently analyzed 
the $K$-band interferometric data on \mbox{$o$~Cet}\ obtained with VLTI/VINCI 
and compared the visibilities observed at 
phases from $\sim$0.1 to $\sim$0.5 with those predicted by dynamical 
model atmospheres in the fundamental mode (including the P and 
M models) as well as in the first overtone mode.   
They show that the observed visibilities can be best explained by 
the dynamical models pulsating in the fundamental mode.  
The Rosseland radius derived from the observed visibilities 
is found to be consistent with the period-radius-relation of the 
fundamental-pulsator models, but not with that of the models 
pulsating in the first overtone mode.  
The analysis of Woodruff et al. (\cite{woodruff04}) further suggests 
the P model as the best-fit model for the VINCI observations of 
\mbox{$o$~Cet}.  Fedele et al. (\cite{fedele05}) have also obtained a 
similar result for another Mira variable, \mbox{R~Leo}, based on the 
analysis of the VINCI data.  These results 
on the pulsation mode of Mira variables are consistent with the 
conclusion drawn by Wood et al. (\cite{wood99}) from 
the comparison of theoretical pulsation models with MACHO 
observations of long period variables in the Large Magellanic 
Cloud as well as with the pulsation velocities derived from line 
profiles (Scholz \& Wood \cite{scholz00}).  

Using the density stratification obtained by a self-excited 
pulsation model at a given cycle and phase, the temperature 
stratification and molecular/atomic abundances in the non-gray
atmosphere are calculated 
with the assumption of local thermodynamical equilibrium (LTE) 
and radiative equilibrium in spherical symmetry, as described 
in Bessell et al. (\cite{bessell96}) and HSW98.  The solar chemical 
composition is used in the calculations of these atmospheric models.  
The next step is to calculate intensity profiles as well as 
synthetic spectra in the 11~\mbox{$\mu$m}\ region 
using the temperature and density stratifications and 
the velocity field of each model.  For the positions and 
the transition probabilities of \mbox{H$_2$O}\ lines, we use the HITEMP 
database (Rothman \cite{rothman97}) 
as well as the line list compiled by PS97, 
which is deemed to be the most 
extensive line list of \mbox{H$_2$O}\ to date.  The line 
shape in the rest frame 
is assumed to be a Voigt profile with a micro-turbulent 
velocity of 5~\mbox{km s$^{-1}$}, which is close to the value derived 
for the Mira variable R~Leo by Hinkle \& Barnes (\cite{hinkle79}).  
The monochromatic intensity profile is calculated by 
evaluating the intensity at multiple impact parameters 
with respect to the stellar disk in the observer's frame 
(e.g., Mihalas \cite{mihalas78}), 
with the radial velocity of each layer taken into account.  
The monochromatic visibility can be obtained by 
taking the Hankel transform (two-dimensional Fourier transform 
in a centrosymmetric case) of the monochromatic intensity profile.  
In the calculation of visibilities, we adopt a distance 
of 107~pc to \mbox{$o$~Cet}\ based on the HIPPARCOS parallax 
(Knapp et al. \cite{knapp03}; see also the discussion of ISTW04 
about the distance of \mbox{$o$~Cet}).

The monochromatic intensity profiles as well as the monochromatic 
visibilities calculated in the relevant wavelength region are then 
spectrally convolved with a response function, which represents the
bandpass used in the ISI observations.  
As described in Ohnaka (\cite{ohnaka04a}, \cite{ohnaka04b}), 
we approximate the response function with a top-hat function 
with a width of 0.17~\mbox{cm$^{-1}$}.  When comparing model visibilities 
calculated in this manner with those observed, it is necessary to take
the flux contribution of an extended dust shell into account.  The presence
of such a dust shell appears as a steep visibility drop at low spatial 
frequencies and lowers the visibility resulting from the 
underlying atmosphere by an amount equal to the fractional flux 
contribution of the dust shell.  Therefore, the total visibility 
$V_{\omega}^{\rm total}$, 
which can be compared with the ISI observations, is calculated as 
\begin{equation}
\label{vis_dilute}
  V_{\omega}^{\rm total} = (1 - f_{\rm dust}) V_{\omega}^{\rm atm} \, ,
\end{equation}
where $f_{\rm dust}$ is a constant representing the fraction of the 
continuous dust emission from the dust shell in the 11~\mbox{$\mu$m}\ 
region, and $V_{\omega}^{\rm atm}$ is the (spectrally convolved) 
visibility predicted by the atmospheric models.  
The spectrum (i.e., emergent flux) is obtained by 
integrating the intensity over the stellar disk.  The 
spectrum is then convolved with a Gaussian representing 
the resolution of the spectroscopic instrument and the macro-turbulent 
velocity in the atmosphere of Miras.  The spectral resolution 
of the TEXES instrument corresponds to $\sim$3~\mbox{km s$^{-1}$}, and 
we adopt a macro-turbulent velocity of 3~\mbox{km s$^{-1}$}, as assumed 
in Ohnaka (\cite{ohnaka04a}).  This leads to a Gaussian 
with a FWHM of 4.24~\mbox{km s$^{-1}$}.  In order to take into account 
the flux contribution of the dust shell, which amounts up 
to $\sim$60--70\% in the mid-infrared, 
the convolved and normalized spectrum is diluted 
as follows:
\begin{equation}
\label{spec_dilute}
 F_{\omega}^{\rm diluted} = (1 - f_{\rm dust})F_{\omega} + f_{\rm dust} \, ,
\end{equation}
where $F_{\omega}^{\rm diluted}$ is the final spectrum, 
and $F_{\omega}$ is the spectrum emerging from the atmosphere 
without the presence of a dust shell.

\section{Comparison with the observed spectra and angular diameters}
\label{sect_comp}

\begin{figure*}
\centering
\sidecaption
\includegraphics[width=12cm]{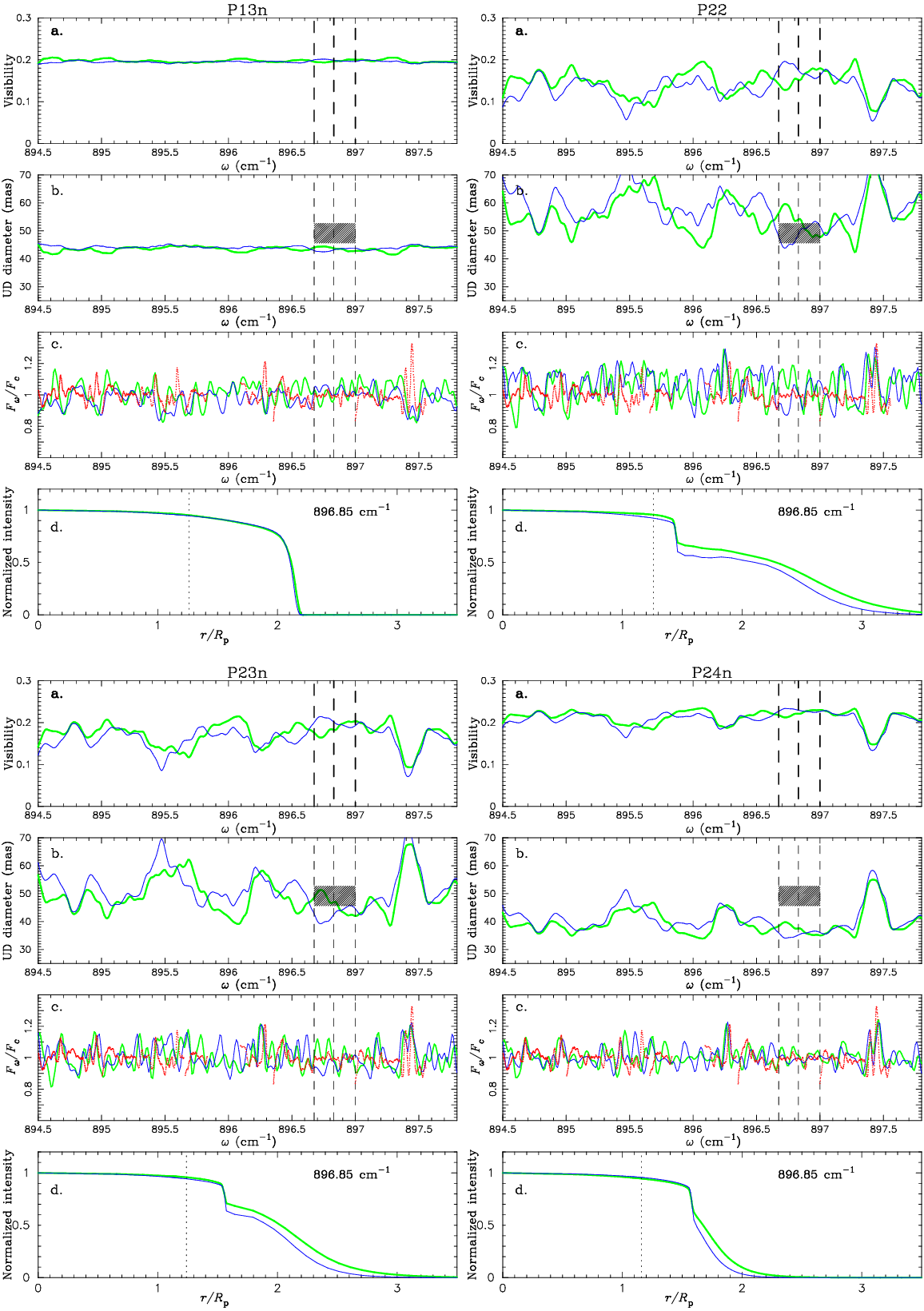}
\caption{Comparison of the mid-infrared interferometric and 
spectroscopic observations with the predictions of selected 
P models.  At each phase, 
the panels {\bf a}, {\bf b}, {\bf c}, and {\bf d} show the predicted 
visibilities, uniform-disk diameters, spectra, and intensity 
profiles at 896.85~\mbox{cm$^{-1}$}, respectively.  
In all panels, the model predictions using the HITEMP database 
and the PS97 line list are represented with the thick (green) and thin 
(blue) solid lines, respectively.  
The visibilities and uniform-disk diameters 
are derived for a projected baseline length of 30~m.  
In the panel {\bf b}, the dashed lines represent the bandpasses 
used in the ISI observations, and the range of the uniform-disk
diameters observed with ISI at relevant phases, from 0.24 to 0.36, is 
represented as the hatched region.  
The (red) dots in the panel {\bf c} show the TEXES spectrum, 
which was read off Fig.~2 of WHT03a. 
The visibilities, uniform-disk diameters, and spectra are 
redshifted by 83~\mbox{km s$^{-1}$}\ to match the observation.  
The intensity profiles in the panel {\bf d} are plotted in units 
of the radius of the parent star (\mbox{$R_{\rm p}$}).  The radii defined by the 
1.04~\mbox{$\mu$m}\ continuum ($R_{1.04}$) are shown by the vertical dotted 
lines.  
}
\label{res_pseries1}
\end{figure*}

WHT03a present ISI observations of \mbox{$o$~Cet}\ 
at phases ranging from 0.99 to 0.36, while the high-resolution
11~\mbox{$\mu$m}\ spectrum was obtained at phase 0.36.  
Therefore, we focus on model calculations for phases close to 0.36, 
at which both spectroscopic and interferometric data are available.  
However, it is necessary to make allowances for the uncertainty 
in determining the phase observationally as well as in assigning 
the observational phase to a model phase of a specific 
cycle, as described in ISTW04.   This uncertainty is of the order of 
$\pm 0.1$, and therefore, we consider models computed for phases 
from 0.2 to 0.5 in different cycles, as listed in
Table~\ref{table_model}.  For each 
model, we calculate the intensity profiles, visibilities, and spectra, 
as described in the previous section.  
The $f_{\rm dust}$ term representing the flux contribution of the dust 
shell was derived by WHT03a from the fitting of the observed 
visibilities with two parameters, the uniform-disk diameter and 
the fraction of the flux coming from the stellar disk (the 
$A$ parameter in WHT03a).  
We adopt $A = 0.3$ for phase 0.36 as given in Table~1 of WHT03a, 
and we use $f_{\rm dust} = 1 - A = 0.7$ in the calculations 
presented below.

\subsection{P series}
\label{subsect_p}

We first test the P models in the zeroth and first cycles.  
It has turned out that the synthetic spectra predicted by the 
P05 model are in fair agreement with the TEXES observation, but the 
uniform-disk diameters predicted by this model within the ISI 
narrow bandpasses are remarkably smaller than the 
observed values.  
On the other hand, 
the P models in the first cycle (P12, P13n, P14n, and P15n) 
predict the uniform-disk diameters marginally in 
agreement with the ISI observations, but cannot reproduce the 
high-resolution 11~\mbox{$\mu$m}\ spectra.  
As an example, the top left panel of Fig.~\ref{res_pseries1} 
shows the comparison between the observations and model predictions 
for the P13n model.  
The visibilities plotted in Fig.~\ref{res_pseries1} were derived 
for a projected baseline length of 30~m, which is approximately the 
mean of the baseline lengths used by W00 and WHT03a.  
The contribution of the extended dust shell is included in these 
model visibilities, as expressed in Eq. (\ref{vis_dilute}).  
The uniform-disk diameters plotted in Fig.~\ref{res_pseries1} were 
derived by fitting the model visibilities predicted for the 
30~m baseline with uniform disks.  Note that the model 
visibilities resulting from the atmosphere alone, excluding the
contribution of the dust shell, are used in the calculation of 
the uniform-disk diameters, because the uniform-disk diameters derived
by WHT03a are already corrected for the flux contribution of 
the dust shell.  
The hatched region in the figure represents 
the range of the uniform-disk diameters measured at 
phases around 0.36, that is, from 0.24 to 0.36, 
making allowances for the uncertainty in the determination of phase 
(note that W00 and WHT03a do not present ISI measurements at phase
later than 0.36).  
The plot for the P13n model in Fig.~\ref{res_pseries1} illustrates 
that the uniform-disk diameters predicted by this model -- whether 
using the HITEMP database or the PS97 line list -- are in rough 
agreement with the observed values, although the model predictions 
are slightly smaller than the observations.  

Comparison between the synthetic spectra and the TEXES observation 
shows that the P13n model can approximately reproduce the absence of 
salient spectral features in the ISI bandpasses, which are marked with 
the dashed lines in the figure (the model spectra obtained by 
Eq. (\ref{spec_dilute}) were scaled to fit the observed spectra).  
It should be noted here that a huge number of \mbox{H$_2$O}\ lines 
mask the true continuum and what appears to be a continuum in the 
TEXES spectrum (for example, the relatively flat portion between 896.5 
and 897.3~\mbox{cm$^{-1}$}) is a result of optically thick emission from the 
\mbox{H$_2$O}\ lines, as discussed in detail below.  
However, this model (and also the other P models in the first cycle) 
cannot reproduce the observed profiles of the strong \mbox{H$_2$O}\ lines 
at $\sim$896.3 and $\sim$897.4~\mbox{cm$^{-1}$}.  
It is worth mentioning that what appears to be absorption (e.g., 
at 897.4 and 897.5~\mbox{cm$^{-1}$}) simply represents the positions where 
the \mbox{H$_2$O}\ optical depth is very 
small (note that the \mbox{H$_2$O}\ optical depth at a given wavenumber 
depends on the velocity field of each model as well).  
And the feature appearing as an emission core at $\sim$897.45~\mbox{cm$^{-1}$}\ 
is in fact the emission of an \mbox{H$_2$O}\ line.  
In any case, the \mbox{H$_2$O}\ line profiles predicted by the P models in the 
first cycle are in disagreement with the observations.

The synthetic spectra calculated with the HITEMP database and 
the PS97 line list show differences in the weak and moderately strong 
\mbox{H$_2$O}\ features.  Since the HITEMP database is an extention of the 
HITRAN database to a temperature of 1000~K, which is still too low 
to deal with temperatures in the atmosphere of Mira variables, 
the differences may be attributed to the incompleteness of high 
excitation lines in the HITEMP database.  
However, while the PS97 line list was generated for a temperature of
4000~K and is regarded as the most extensive \mbox{H$_2$O}\ line list, 
there are also problems with this line list when applied to 
the atmosphere of cool stars.  For example, Jones et al. 
(\cite{jones02}) point out that the use of the PS97 line list 
in the models of M stars 
leads to \mbox{H$_2$O}\ bands that are too strong for a given temperature.  
These differences in 
the \mbox{H$_2$O}\ line lists suggest that the differences between the 
observed and synthetic spectra may at least partially result 
from the uncertainties of the \mbox{H$_2$O}\ line positions and strengths.

Now, we examine the P models in the second and third cycles.  
The results obtained with the P22, P23n, and P24n models 
are shown in Fig.~\ref{res_pseries1}.  
The P22 model predicts uniform-disk 
diameters in agreement with the ISI observations, but the \mbox{H$_2$O}\ lines 
predicted by this model appear too pronounced,  
compared to the TEXES observation.  For example, in the region 
between 896.5 and 897.3~\mbox{cm$^{-1}$}, where the ISI bandpasses are located, 
the model cannot reproduce the observed, continuum-like spectrum.  
On the other hand, the P23n model can 
approximately reproduce the observed high-resolution spectrum 
and the 11~\mbox{$\mu$m}\ uniform-disk diameters simultaneously.  
The synthetic spectra are in fair agreement with the observed, 
continuum-like spectrum (e.g., between 895.0 and 895.4~\mbox{cm$^{-1}$}\ 
as well as between 896.5 and 897.3~\mbox{cm$^{-1}$}), although the model still 
predicts the \mbox{H$_2$O}\ lines to be too pronounced.  
The uniform-disk diameters predicted within the ISI bandpasses 
centered around 896.8~\mbox{cm$^{-1}$}\ are also in reasonable 
agreement with the ISI observations, although they are still 
somewhat smaller than the observed values.  

\begin{figure}
\centering
\resizebox{\hsize}{!}{\includegraphics{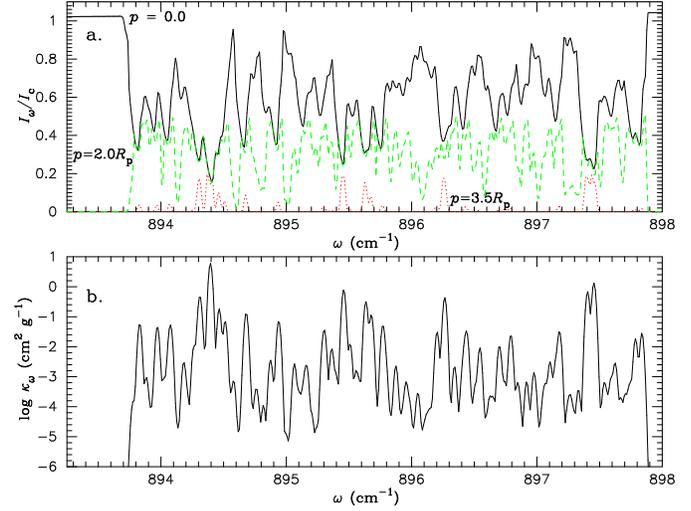}}
\caption{{\bf a:} The spectra predicted by the P23n model at
  different impact parameters.  The spectrum at the stellar disk
  center ($p = 0.0$) is plotted with the solid line, while the 
  spectrum at the extended wing ($p = 3.5$~\mbox{$R_{\rm p}$}, see also 
  Fig.~\ref{intensplot_P23n}) is plotted with the dotted line.  
  The dashed line represents the spectrum between the above two 
  cases, at $p = 2.0$~\mbox{$R_{\rm p}$}.  The contribution of the dust emission 
  is not included in the spectra. 
  {\bf b:} The opacity of \mbox{H$_2$O}\ lines in an upper layer of the P23n 
  model ($T = 1020$~K and $\log P_{g} = -2.4$~dyn~cm$^{-2}$).  
  The spectra and the opacity are calculated with the HITEMP 
  database.  They are redshifted by 83~\mbox{km s$^{-1}$}\ to match the observation 
  and not convolved with a Gaussian representing the 
  instrumental as well as macro-turbulent broadening.  
}
\label{intens_sp_P23n}
\end{figure}

The flat appearance of the spectrum, despite the presence of a number 
of \mbox{H$_2$O}\ lines, is a result of the filling-in effect, as 
proposed by Ohnaka (\cite{ohnaka04a}) to explain the mid-infrared
spectra and angular sizes of Mira variables based on a semi-empirical 
model.  
This effect is illustrated in Fig.~\ref{intens_sp_P23n}, where 
the synthetic spectra at three different impact parameters ($p$) are 
plotted.  The spectrum predicted at the disk center (solid line) 
exhibits prominent absorption features, while the spectra 
predicted in the extended wing (dashed and dotted lines) show emission 
features at the 
positions of \mbox{H$_2$O}\ lines.  It should be noted that the spectrum 
predicted at $p = 2.0$~\mbox{$R_{\rm p}$}\ shows pronounced emission at 
wavenumbers where the spectrum at the disk center shows absorption 
features (e.g., 894.7, 895.15, and 896.7~\mbox{cm$^{-1}$}).  And what appears 
to be absorption features in the spectrum at $p = 2.0$~\mbox{$R_{\rm p}$}\ 
(e.g., 894.1, 894.6, 894.8, and 895.0~\mbox{cm$^{-1}$}) simply represents 
the positions where the \mbox{H$_2$O}\ line opacity is very low, as can be 
seen in Fig.~\ref{intens_sp_P23n}b.  This emission 
from the extended \mbox{H$_2$O}\ layers fills in the absorption expected 
at $p \la \mbox{$R_{1.04}$}$, leading to a featureless spectrum.

\begin{figure}
\centering
\resizebox{\hsize}{!}{\includegraphics{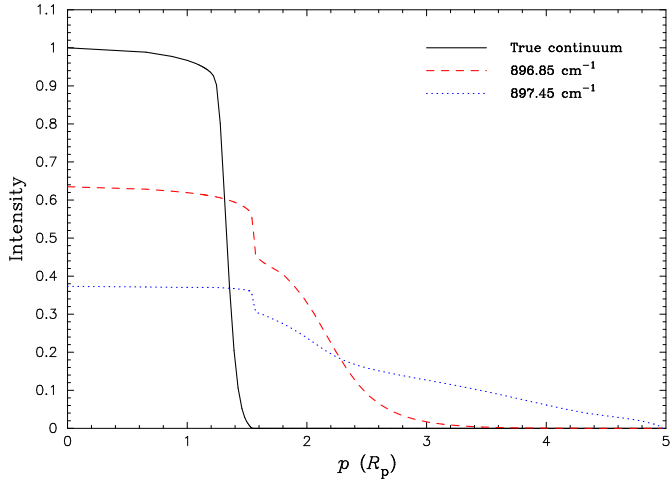}}
\caption{The intensity profiles in the true continuum (solid line), 
in the pseudo-continuum at 896.85~\mbox{cm$^{-1}$}\ (dashed line), 
and in a strong \mbox{H$_2$O}\ line at 897.45~\mbox{cm$^{-1}$}\ (dotted line) 
predicted by the P23n model.  The intensity profiles 
are calculated with the HITEMP database and spectrally convolved to
match the spectral resolution of the ISI measurements. 
The profiles are normalized with the value at $p = 0$ of the 
intensity in the true continuum.  
}
\label{intensplot_P23n}
\end{figure}

The P23n model can also approximately predict the inverse P-Cyg 
profiles of strong \mbox{H$_2$O}\ lines observed at $\sim$894.9, 895.4, and 
897.4~\mbox{cm$^{-1}$}, but there are still slight 
discrepancies in the line depths and positions.  
The model predicts the line profiles to be broader than those observed, 
and this appears to imply that the micro- and macro-turbulent velocities 
used in the calculation may be too large.  
However, while the use of smaller micro- and macro-turbulent
velocities can improve the agreement with the observed line profiles 
of these strong \mbox{H$_2$O}\ lines, 
it also makes the weak and moderately strong \mbox{H$_2$O}\ lines much more 
conspicuous in the ISI bandpasses.  Therefore, the difference 
in the line profiles may result from the uncertainties of the 
\mbox{H$_2$O}\ line data as well as the uncertainties of the 
atmospheric structure and/or the concept of constant, isotropic 
micro- and macro-turbulence itself.   
Figure~\ref{intensplot_P23n} shows the intensity profiles predicted 
by the P23n model at $\sim$897.45~\mbox{cm$^{-1}$}\ (on a strong \mbox{H$_2$O}\ line), 
896.85~\mbox{cm$^{-1}$}\ (in the ``pseudo-continuum''), and in the true 
continuum, which was calculated with the \mbox{H$_2$O}\ line opacity set to 
zero.  
The intensity profile in the strong 
\mbox{H$_2$O}\ line (dotted line) shows a very extended wing with rather high 
intensities as compared to that in the pseudo-continuum (dashed line), 
which makes the line appear in emission above the pseudo-continuum.  

\begin{figure}
\centering
\resizebox{\hsize}{!}{\includegraphics{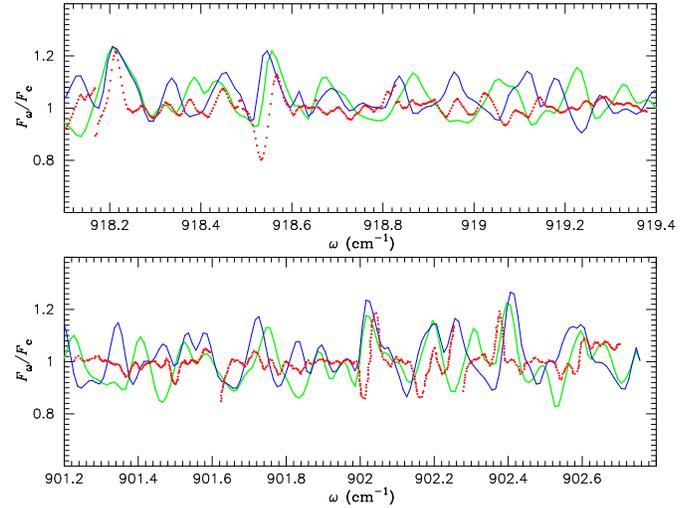}}
\caption{The synthetic spectra calculated with the P23n model are
  represented with the thick (HITEMP database) and thin (PS97 line list)
  lines. The observed spectrum is plotted with the dots. 
  The synthetic spectra are redshifted by 83~\mbox{km s$^{-1}$}\ to match the
  observation and already convolved with a Gaussian representing the 
  macro-turbulence in the stellar atmosphere and the 
  instrumental broadening.  
}
\label{spectra2_P23n}
\end{figure}

Figure~\ref{spectra2_P23n} shows a comparison between the observed 
and synthetic spectra calculated with the P23n model in another 
wavelength region presented in WHT03a.  
The quality of the agreement with the observation is fair, 
though not fully satisfactory.  The rather featureless spectrum observed 
in the region between 918.6 and 919.4~\mbox{cm$^{-1}$}\ as well as 
between 901.2 and 902~\mbox{cm$^{-1}$}\ is reproduced by the model to some 
extent, though the synthetic spectra tend to show too prominent 
spectral features.  The strengths of the prominent \mbox{H$_2$O}\ lines 
at 918.2, 918.55, 902.0, and 902.37~\mbox{cm$^{-1}$}\ are also in rough agreement, 
but there are still discrepancies between the observed line profiles 
and those predicted by the model.  

\begin{figure*}
\sidecaption
\includegraphics[width=12cm]{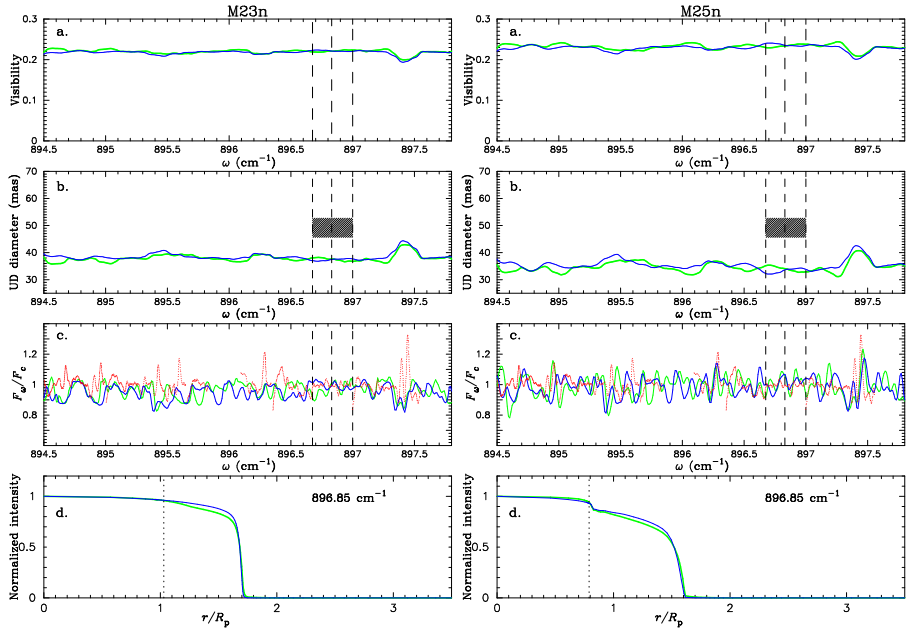}
\caption{
Comparison of the mid-infrared interferometric and 
spectroscopic observations with the predictions of selected 
M models.  See also the caption to Fig.~\ref{res_pseries1}.  
}
\label{res_mseries1}
\end{figure*}

We found out that the other P models at phases later than 2.3 
cannot reproduce the observed spectrum and uniform-disk diameters 
simultaneously: the 11~\mbox{$\mu$m}\ uniform-disk diameters are 
too small compared to the ISI measurements.  
The result obtained with the P24n model is 
shown in Fig.~\ref{res_pseries1} (lower right) as an example.  Obviously, 
this model predicts the 11~\mbox{$\mu$m}\ uniform-disk diameters to be 
considerably smaller than the observed values, while the synthetic 
spectra are in reasonable agreement with the observed spectrum.  
This results from the compactness of the atmospheres of the P models 
at phases later than 2.3, as compared to that of the P23n model.

\subsection{M series}
\label{subsect_m}

In this subsection, we discuss the spectra and 
the uniform-disk diameters predicted by the M models, 
which have a higher mass (1.2~\mbox{$M_{\sun}$}) than the P models (1.0~\mbox{$M_{\sun}$}).  
Figure~\ref{res_mseries1} shows a comparison of the spectra 
and uniform-disk diameters calculated with the M23n and M25n 
models as examples.  
The visibilities and uniform-disk diameters plotted in the figure 
are derived for a projected baseline length of 30~m, as 
described in the previous subsection.  
The figure reveals that the M23n model can reproduce neither the 
observed 11~\mbox{$\mu$m}\ uniform-disk diameters nor the 11~\mbox{$\mu$m}\ 
spectrum.  The predicted uniform-disk diameters are smaller than 
the ISI measurements, and the profiles of the strong \mbox{H$_2$O}\ lines 
(e.g., at 897.4~\mbox{cm$^{-1}$}) are in clear disagreement with the observation.  
The synthetic spectrum predicted by the M25n model is in 
rough agreement with the observation in the continuum-like part within 
the ISI bandpasses as well as in the profiles of the strong \mbox{H$_2$O}\
lines (for example, the lines at 897.4~\mbox{cm$^{-1}$}).  
However, the uniform-disk diameters calculated 
with this model are also much smaller than the observed values, 
as in the case of the M23n model.  
We note here, however, that the strong lines are formed in the 
uppermost layers and may be moderately affected by the fact that 
the model atmospheres are cut off at $5 \times \mbox{$R_{\rm p}$}$ (see HSW98), 
and the transition to the circumstellar envelope is not taken into 
account.  

It turned out that none of the M models considered here 
can simultaneously reproduce the observed 11~\mbox{$\mu$m}\ spectra 
and uniform-disk diameters.   The atmospheres of the M models 
calculated for phases used in the present work 
are considerably more compact compared to those of the P models, which results 
in uniform-disk diameters much smaller than the observed
values.  

\section{Discussion}

The comparison of the model spectra and uniform-disk diameters 
with the TEXES and ISI observations discussed in the previous section
demonstrates 
that the P23n model is the most appropriate one to explain the 
observations at phase 0.36.  
However, even this ``best'' model cannot 
provide fully satisfactory agreement with the observational results: 
the predicted uniform-disk diameters tend to be somewhat smaller than
the observed values, and the synthetic spectra show too 
pronounced \mbox{H$_2$O}\ features.  

One possibility to reconcile this 
discrepancy is to assume a flux contribution from the dust 
shell higher than that derived by WHT03a.  
They fitted the observed visibilities with two parameters, that is, 
the uniform-disk diameter and the fractional flux contribution of 
the stellar disk.  However, as the intensity profiles plotted in 
Figs.~\ref{res_pseries1} and \ref{intensplot_P23n} illustrate, 
uniform disks are not necessarily an appropriate approximation to
characterize 
the real intensity distributions of Mira variables, and therefore, 
the dust flux contribution derived from the two-parameter fit 
may be affected by this assumption, even if the formal error of 
the fit is small.  

\begin{figure}
\centering
\resizebox{\hsize}{!}{\includegraphics{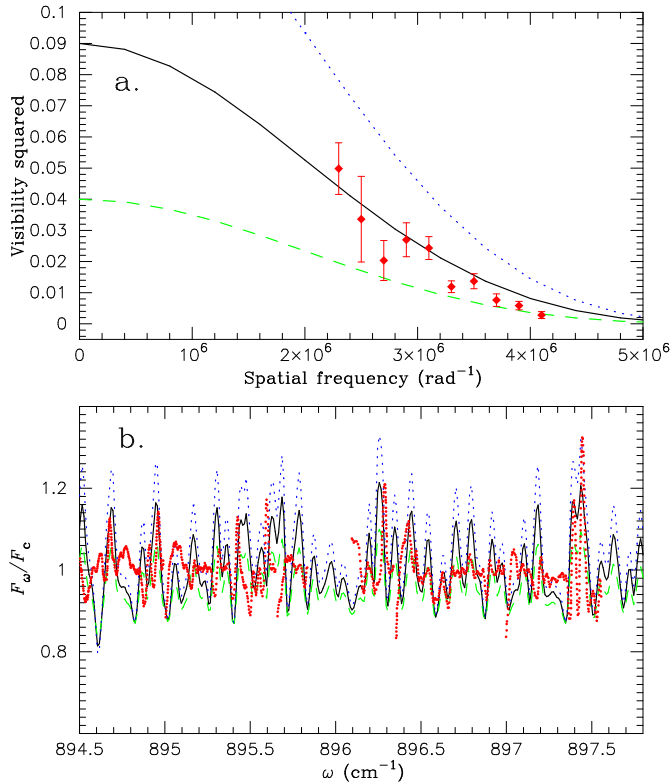}}
\caption{Comparison of the observed visibilities and spectra 
with those calculated with the P23n model using three different 
values for the fractional flux contribution of the dust shell.  
In both panels, the dotted, solid, and dashed lines represent 
the calculations with $f_{\rm dust} = 0.6$, 0.7, and 0.8, 
respectively.  
{\bf a.} The filled diamonds represent the 
ISI visibility points obtained at phase 0.36 (2001 Dec. 19).  
{\bf b.} The 11~\mbox{$\mu$m}\ spectrum obtained with the TEXES 
instrument at phase 0.36 is plotted with the (red) dots.  
}
\label{visfreq_P23n}
\end{figure}

We check the influence of the uncertainty of 
the flux contribution of the dust shell on the synthetic spectra 
and visibilities by changing $f_{\rm dust}$, the term representing the 
flux contribution of the dust shell.  
Figure~\ref{visfreq_P23n} shows the result of 
calculations with the P23n model using three different values 
for $f_{\rm dust}$: 0.6, 0.7, and 0.8.  The visibilities predicted 
at 896.85~\mbox{cm$^{-1}$}\ are plotted as a function of spatial frequency, 
together with the ISI visibilities observed within the bandpass centered 
around 896.85~\mbox{cm$^{-1}$}.  These ISI visibilities were obtained by 
binning the original calibrated visibilities of \mbox{$o$~Cet}\ measured on 
2001 Dec. 19 at phase 0.36 (J.~Weiner, M.~J.~Ireland, priv. comm.).  
The figure 
suggests that a flux contribution of the dust shell slightly higher 
than 0.7 can explain the observed visibilities.  
As the lower panel of the figure shows, an increase of the dust flux 
makes the \mbox{H$_2$O}\ features less pronounced, resulting in better 
agreement with the observed, continuum-like spectra.   
However, it also renders the strong \mbox{H$_2$O}\ lines (e.g., at
897.4~\mbox{cm$^{-1}$}) weaker, resulting in a poorer match with the 
observed line profiles.  Thus, while the uncertainty of the flux 
contribution of the dust shell can be partially responsible 
for the discrepancy between the observations and the model, 
a mere change of $f_{\rm dust}$ is unlikely to solve the problem.  

Certainly, dynamical model atmospheres currently available 
cannot yet explain all observational aspects of Mira variables.  
A discrepancy between theory and observation can also be seen 
in the temporal variation of the spectral features.  
Weiner (\cite{weiner04}) presents high-resolution 11~\mbox{$\mu$m}\ 
spectra obtained at two different phases, 0.15 and 0.47, 
in the same cycle (see Fig.~4 in Weiner \cite{weiner04}).  
These spectra, which cover approximately the same wavelength range 
as we study here, exhibit only modest temporal variations, 
while the models predict much more pronounced changes between 
phase 2.2 and 2.4, as can be seen in Fig.~\ref{res_pseries1}.  

It should be stressed here that there are numerous uncertainties
of model parameters and modeling assumptions that may readily prevent 
good agreement between the observations and model predictions. 
For example, the only well-known stellar parameter of \mbox{$o$~Cet}\ is its 
period.  
The luminosity (and hence the effective temperature as well)
derived from the observed bolometric flux and the star's distance 
agrees with the model value for the Mira star R~Leo, if the HIPPARCOS 
distance of Knapp et al. (\cite{knapp03}) is adopted (82 pc), 
whereas it is higher than the model value in the case of \mbox{$o$~Cet}\ 
(107~pc; see the discussion in ISW04, ISTW04, and 
Woodruff et al. \cite{woodruff04}). 
The uncertainty of the distance also directly affects the uniform-disk 
diameters predicted by the models.  
The mass of \mbox{$o$~Cet}\ (``around 1~\mbox{$M_{\sun}$}'', Wyatt \& Cahn \cite{wyatt83}) 
and its metallicity are only a reasonable guess.  Dynamical models
covering a wider parameter range are being constructed but not yet 
available.  
Also, there are uncertainties in the treatment of physical 
processes included in the calculations of self-excited pulsation 
models and dynamical model atmospheres.  Self-excited pulsation models
may be moderately influenced by the treatment of convection.  
The numerous approximations adopted 
in the calculation of the time-dependent structure of the non-gray 
atmosphere as described in HSW98 may also be responsible for 
the discrepancy between the observations and model predictions.  
Note also that the \mbox{H$_2$O}\ line calculations presented here are 
performed with the assumption of LTE.  Noticeable deviations from LTE 
may occur in the uppermost layers with low densities and affect 
particularly strong lines.   
For another molecule with strong features, TiO, M.~J.~Ireland et al. 
(in preparation) suggest that the strong TiO absorption bands are 
significantly affected by non-LTE effects.

One of the physical processes of great importance in understanding 
the structure of the outer atmosphere of Mira variables is 
the dust formation.  Jeong et al. (\cite{jeong03}) present a 
detailed time-dependent hydrodynamic calculation for the oxygen-rich 
Mira variable IRC$-$20197 (IW~Hya), which shows that the dust formation 
can take place at 3--5~\mbox{$R_{\star}$}.  Although the parameters of 
IRC$-$20197 (lower effective temperature, higher luminosity, 
and higher mass loss rate compared to \mbox{$o$~Cet}) suggest that this object is
much more evolved than \mbox{$o$~Cet}, the dust formation at such a 
vicinity of the star can also be expected in \mbox{$o$~Cet}.  
The formation of dust grains would have two effects on 
the comparison with the spectra and the angular size observed in the 
mid-infrared.  Firstly, as discussed above, the flux contribution 
of the dust shell renders \mbox{H$_2$O}\ spectral features less pronounced.  
The derivation of uniform-disk diameters from visibilities measured at 
high spatial frequencies can also be affected for the following 
reason.  That is, while the dust flux contribution appears as a steep 
visibility drop at low spatial frequencies if the dust shell lies 
well above the \mbox{H$_2$O}\ line-forming layers of the atmosphere, 
this is not the case if dust forms 
very close to the star and the dust shell is not well separated from 
the layers where \mbox{H$_2$O}\ lines form.  
In this case, the visibility function shows a more complex shape 
at low spatial frequencies (e.g., Lopez et al. \cite{lopez97}), 
which makes it difficult to derive 
reliably the uniform-disk diameters of the underlying atmosphere.  
Secondly, the dust formation in the upper 
layers can change the temperature and density stratifications 
due to the backwarming effect 
as well as dynamical effects resulting from the onset of mass
outflows (e.g., H\"ofner et al. \cite{hoefner96}; 
Bedding et al. \cite{bedding01}; Jeong et al. \cite{jeong03}).  
In fact, a recent attempt to model the formation of dust grains and 
their effects on the atmospheric structure in the upper layers 
of the P and M models confirms that the dust formation does take 
place in those upper layers, but it also reveals serious problems 
in putting observational constraints on the details of the dust
formation process (M.~J.~Ireland et al., in preparation).

\section{Concluding remarks}

We have presented a comparison of 
the mid-infrared spectra and angular sizes predicted by dynamical model 
atmospheres with the results of the TEXES and ISI observations.  
The P model in the second cycle calculated for a phase close to the 
observations can fairly reproduce the observed mid-infrared 
uniform-disk diameters and high-resolution spectra without 
introducing ad-hoc components.   
This result lends support to the picture obtained 
by semi-empirical models: the observed, continuum-like 11~\mbox{$\mu$m}\ 
spectra are the result of the filling-in effect due to the emission from 
extended water vapor layers, while these water vapor layers manifest 
themselves as an increase of the angular size from the near-infrared 
to the mid-infrared.  On the other hand, the M models considered 
in the present work fail to 
reproduce the observed 11~\mbox{$\mu$m}\ uniform-disk diameters 
and the line profiles of prominent \mbox{H$_2$O}\ features, which can 
be attributed to their rather compact atmospheres at phases 
considered here.   

Still, the ``best'' model in the P series (P23n) predicts the
uniform-disk diameters at 11~\mbox{$\mu$m}\ 
to be somewhat smaller than the observed values.  The predicted \mbox{H$_2$O}\ 
spectral features appear to be too pronounced and show too significant
temporal variations compared to the observed spectra.  
These discrepancies may be attributed to the uncertainties of the 
input parameters such as the uncertainties of \mbox{H$_2$O}\ line data 
(line positions and line strengths), the unknown stellar mass and/or 
luminosity, as well as those of the treatment of relevant 
physical processes.  
From the density and temperature conditions in the 
extended configurations of the P and M models, 
we also suspect that dust formation may occur in the upper atmospheric 
layers.  
More refined dynamical models (with dust formation) would be 
needed to obtain a definitive explanation for the increase of 
the angular diameter from the near-infrared to the mid-infrared as
well as the flat mid-infrared spectra observed in Mira variables.

\begin{acknowledgement}
We would like to thank J.~Weiner and M.~J.~Ireland for kindly 
providing us with the ISI data of \mbox{$o$~Cet}.  
This research was in part (M.~Scholz and P.~R.~Wood) supported by the
Australian Research Council and the Deutsche Forschungs\-gemeinschaft 
within the linkage project ``Red Giants'', and by a grant of the 
Deutsche Forschungs\-gemeinschaft on ``Time dependence of Mira 
Atmospheres''.  

\end{acknowledgement}

\end{document}